\documentclass[a4paper]{jpconf}
\usepackage{graphicx}
\begin{document}
\title{UrQMD calculations of two-pion HBT correlations in p+p and Pb+Pb collisions at LHC energies}

\author{Qingfeng Li}

\address{School of Science, Huzhou Teachers College, Huzhou 313000,
P.\ R.\ China}

\ead{liqf@hutc.zj.cn}

\author{Gunnar Gr\"af and Marcus Bleicher}

\address{Frankfurt Institute for Advanced Studies (FIAS), Johann Wolfgang Goethe-Universit\"{a}t, Ruth-Moufang-Str.\ 1, D-60438 Frankfurt am Main, Germany}

\begin{abstract}
Two-pion Hanbury-Brown-Twiss (HBT) correlations for p+p and central Pb+Pb
collisions at the Large-Hadron-Collider (LHC) energies are investigated with
the ultra-relativistic quantum
molecular dynamics model combined with a correlation afterburner. The transverse momentum dependence of the
Pratt-Bertsch HBT radii $R_{long}$, $R_{out}$, and $R_{side}$ is extracted from a three-dimensional
Gaussian fit to the correlator in the longitudinal co-moving system. In the p+p case, the dependence of correlations on the charged particle multiplicity and formation time is explored and the data allows to constrain the formation time in the string fragmentation to $\tau_f \leq 0.8$ fm/c. In the Pb+Pb case, it is found that $R_{out}$ is overpredicted by nearly 50\%. The LHC results
are also compared to data from the STAR experiment at RHIC. For both
energies we find that the calculated $R_{out}/R_{side}$ ratio is always
larger than data, indicating that the emission in the model is less
explosive than observed in the data.
\end{abstract}

\section{Introduction}
The properties of strongly interacting matter are described by the theory of
Quantum-Chromo-Dynamics (QCD). To explore the details of QCD matter under
extreme conditions, one needs to compress and heat up QCD matter to regimes
present microseconds after the Big Bang. Today these conditions can only be
found in the interior of neutron stars or created in heavy-ion collisions at
relativistic energies. Over the last decade the experimental programs at the
Super Proton Synchrotron (SPS) and at the Relativistic Heavy Ion Collider (RHIC) have provided exciting pioneering data on the
equation of state (EoS), the transport properties of the matter created and its
spatial distributions
\cite{Adams:2004yc,Lisa:2000no,Alt:2007uj,Afanasiev:2002mx,Adamova:2002wi,Abelev:2009tp,Back:2004ug,Back:2005hs,Back:2002wb,Abelev:2008ez}. In addition, the Large
Hadron Collider (LHC) at CERN had been designed,  installed, tested,
and repaired in the past two decades and finally, started normal
operation in the end of the year 2009. Since then, a tremendous
amount of experimental data in various aspects of high energy
physics has been obtained and received much attention by theoretical
physicists. And, the extracted bulk properties of the high
temperature fireball created in such ultra-relativistic collisions
have provided unprecedented information for fundamental
investigations of the phase diagram of QCD.
Here we want to explore the expansion properties of the created
matter by investigating the spatial shape of the fireball. Although
it is known that one can not measure the emission time pattern
and/or the spatial profile of the source directly, a
well-established technique, called ``femtoscopy'' or ``HBT'' (see
e.g. \cite{Lisa:2005dd} and references therein) can be employed to
obtain this information. Femtoscopy has been extensively used in the
heavy ion community since it provides the most direct link to the
lifetime and size of nuclear reactions. The ALICE collaboration has
published first results of two-pion Bose-Einstein correlations  in
both p-p \cite{Aamodt:2011kd} and central Pb-Pb \cite{Aamodt:2011mr}
collisions at LHC energies in the beginning of the year 2011. These
experimental results have attracted the research interest of several
theoretical groups
\cite{Aamodt:2011mr,Karpenko:2011qn,Humanic:2010su,Werner:2011fd},
whose models are based on hydrodynamic/hydrokinetic and hadronic
microscopic approaches.

It is often argued that the particle emitting system in p+p collisions is too small to create a medium that exhibits
bulk properties, this is different at a center of mass energy of $\sqrt{s}$= 7 TeV. Here, the particle multiplicity is about the same as in nucleus-nucleus collisions, studied at RHIC. These data suggest that space momentum
correlations are developed even in p+p collisions as soon as high particle multiplicities are achieved. Thus it is worthwhile studying the dependence
of HBT observables on the event multiplicity. As the system created in p+p collisions is still small, an essential quantity
that influences the particle freezeout radii is the formation time in flux tube fragmentation. The formation time sets the scale for a minimum value of the source lifetime - of course followed by resonance decay and rescattering. In this paper we suggest that the recent LHC data on p+p collisions
allow us to determine the formation time in the flux tube break-up.

In this paper we show results for the HBT radii
of two-pion correlations from p+p and central ($<5\%$ of the total cross
section $\sigma_T$) Pb+Pb collisions at LHC energies
$\sqrt{s_{NN}}=7$ and $2.76$ TeV, respectively, from the ultra-relativistic quantum
molecular dynamics (UrQMD) model
\cite{Bass:1998ca,Bleicher:1999xi,Petersen:2008kb,Petersen:2008dd}.
The calculations are compared to ALICE data as well as to those at
the RHIC energy $\sqrt{s_{NN}}=200$ GeV.

The paper is arranged as follows: In Section 2, a brief description
of the UrQMD model and the treatment of the HBT correlations as well
as the corresponding Gaussian fitting procedure are shown. Section 3
gives the main results of the model calculations for p+p and Pb+Pb collisions at LHC energies. Finally, in
Section 4, a summary is given.

\section{Brief description of the UrQMD model and the HBT Gaussian fitting procedure}
UrQMD \cite{Petersen:2008kb,Petersen:2008dd,Li:2011aa} is a microscopic many-body approach to p+p, p+A, and
A+A interactions at energies ranging from SIS up to LHC. It is based on the
covariant propagation of mesons and baryons. Furthermore, it includes rescattering of particles, the
excitation and fragmentation of color strings, and the formation and decay of hadronic resonances. At
LHC, the inclusion of hard partonic interactions in the initial stage is important and is treated via
the PYTHIA \cite{Sjostrand:2006za} model.

Formation time denotes the time it takes for a hadron to be produced from a fragmenting string. A very
common model to describe such a flux tube fragmentation is the Lund string model \cite{Andersson:1983ia}. In the Lund model the formation time consists of the time it takes to produce an quark-antiquark pair and the time it takes for that pair to form a hadron. For the Lund model, both of these times are proportional to the transverse mass of the created hadron and inversely proportional to the string tension. For simplicity, UrQMD uses a constant formation time of $\tau_f=0.8$ fm/c for hard collisions. Since HBT probes the freezeout distribution, the extracted radii are sensitive to the value of $\tau_f$ in small systems at high collision energies where the majority of particles are produced from flux tube fragmentaion.

In the present study, the cascade mode of the latest version (v3.3)
of UrQMD is used (for details of version 3.3. see
\cite{Petersen:2008kb,Petersen:2008dd}). Some predictions and
comparison works with data from reactions at LHC have already been
pursued based on this version and showed encouraging results for the
bulk properties \cite{Mitrovski:2008hb,Petersen:2011sb}.
After the UrQMD simulation, all particles with their phase space
distributions at their respective freeze-out time (last collisions)
are put into an analyzing program using the formalism of
the well known ``correlation after-burner'' (CRAB)
\cite{Pratt:1994uf}. At last, the constructed two-pion HBT
correlator (regardless of charge) in the longitudinally co-moving
system (LCMS) \cite{Pratt:1986cc,Bertsch:1988db} without influence
of residual interactions is fitted (using the $\chi^2$ method) with
a three-dimensional Gaussian form expressed as

\begin{equation}
C(\mathbf{q},\mathbf{k})=1+\lambda(\mathbf{k})
{\rm exp}\left [- \sum_{i,j=out,side,long} q_i q_j R_{ij}^2(\mathbf{k}) \right ]. \label{fit1}
\end{equation}
In Eq.~(\ref{fit1}), $\lambda$ represents the fraction of correlated
pairs \cite{Lisa:2005dd} and $q_i$ is the pair relative momentum
$\mathbf{q}=\mathbf{p}_1-\mathbf{p}_2$ in the $i$ direction, $p_i$
being the momenta of the particles. $long$ is the longitudinal
direction along the beam axis, $out$ is the outward direction along the
transverse component of the average momentum $\mathbf{k}$ of two
particles ($\mathbf{k}_T=|\mathbf{p}_{1T}+\mathbf{p}_{2T}|/2$) and
$side$ is the sideward direction perpendicular to the afore mentioned
directions. The effect of cross terms  with $i\neq j$ on the HBT
radii is found to be negligible in the present fits when the
pseudorapidity cuts $|\eta|<1.2$ and $|\eta|<0.8$ are used for p+p and Pb+Pb cases, respectively, as in experiments, and is
not discussed in this paper.

For central collisions, the HBT radii are, except for an implicit
$\mathbf{k}_T$ dependence, related to regions of homogeneity by
\cite{Wiedemann:1999qn}

\begin{equation}
R_{out}^2 =  \left < (x-\beta_T t)^2 \right >  =  \left < x^2 \right > - 2 \left < \beta_T t x \right >  + \left <\beta_T^2 t^2 \right > , \label{eqn:RO}
\end{equation}

\begin{equation}
R_{side}^2 =  \left < y^2 \right > , \label{eqn:RS}
\end{equation}
and
\begin{equation}
R_{long}^2 =  \left < (z-\beta_L t)^2 \right >  =  \left < z^2 \right > - 2 \left < \beta_L t z \right > + \left < \beta_L^2 t^2 \right > , \label{eqn:RL}
\end{equation}
where $x$, $y$, $z$ and $t$ are the spatio-temporal separation of
the particles in a pair and {\boldmath$\beta$}$ = \mathbf{k}/k_0$.
If no space-momentum correlations are present the regions of
homogeneity and the source size coincide. In central collisions the
relation $\left < x^2 \right > \simeq \left < y^2 \right >$ is
satisfied. Thus $R_{out}^2$ and $R_{side}^2$ differ mainly in the last two
terms of Eq.~(\ref{eqn:RO}). The first of these two terms is
dependent on the strength of the correlation of emission time and
transverse emission position, while the second one is especially
sensitive to the particle emission duration.

For the p+p case, the analysis is done differentially for the different event multiplicities listed in table \ref{Tab:Multiplicities}. Generally, the average $dN_{ch}/d\eta$ from UrQMD is 15$\%$ smaller than the one measured by the ALICE collaboration, in the same charged multiplicity classes.

   \begin{table}
      \begin{center}
      	 \caption{Table of the investigated multiplicity intervals. The first column shows the
		  interval boundaries, the second column the mean charged particle multiplicity per unit
		  of pseudorapidity ($dN_{ch}/d\eta$) from events with at least one charged particle in $|\eta| < 1.2$ from
		  UrQMD. The third column shows the same quantity from ALICE data \cite{Aamodt:2011kd}.}
	 \begin{tabular}{ccc}
       \br
	   $N_{charged}^{|\eta|<1.2}$   & UrQMD $\left < dN_{ch}/d\eta \right >_{|\eta|<1.2}$ & ALICE $\left < dN_{ch}/d\eta \right >_{|\eta|<1.2}$\\
	   \mr
	   1-11   &  2.52    & 3.2  \\
	   12-16  &  5.74    & 7.4  \\
	   17-22  &  8.01    & 10.4 \\
	   23-29  &  10.72   & 13.6 \\
	   30-36  &  13.65   & 17.1 \\
	   37-44  &  16.74   & 20.2 \\
	   45-57  &  20.94   & 24.2 \\
	   58-149 &  27.57   & 31.1 \\
       \br
	 \end{tabular}
	 \label{Tab:Multiplicities}
      \end{center}
   \end{table}

\section{Results}

\subsection{p+p collisions at $\sqrt{s_{NN}}=7$ TeV}
\begin{figure}[!ht]
      \includegraphics[width=0.9\textwidth]{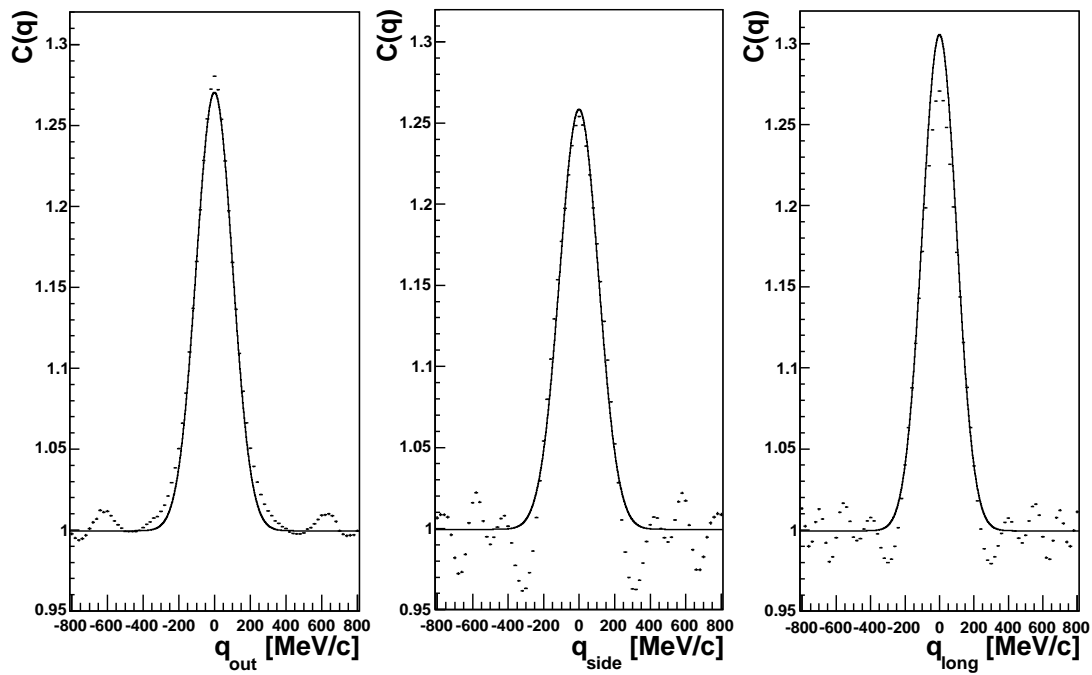}
      \caption{The dots represent projections of the three-dimensional correlation function for
               $k_T$=0.3-0.4 GeV and $N_{charged}^{|\eta|<1.2}$= 23-29. The lines represent a $\chi ^2$ fit of
               Eq. (\ref{fit1}) to the correlation function. Both the result of the fit
               and the correlation function are integrated over a range of $q_i=\pm 0.17$ GeV
               in the other directions for the purpose of projection.
          \label{Fig:CorrFunction}}
   \end{figure}

In this subsection, the results on HBT radii in p+p collisions from the UrQMD model are calculated and compared to ALICE data
\cite{Aamodt:2011kd}. In Fig. \ref{Fig:CorrFunction} the projections in the out, side and long directions of the 3D correlation function together with a projection of the fit for $k_T=|\mathbf{k}_T|=0.3-0.4$ GeV and $N_{charged}^{|\eta|<1.2}$= 23-29 are shown as an example. It can be seen that the calculated correlation function (shown as dots) is well described by a Gaussian fit (lines). However, oscillations of the correlation function are present at larger {\it q}. This indicates that there is a non-gaussian component in the underlying separation distribution of the pion freezeout points. Here we will not investigate this interesting question further.

\begin{figure}[!ht]
     \includegraphics[width=0.9\textwidth]{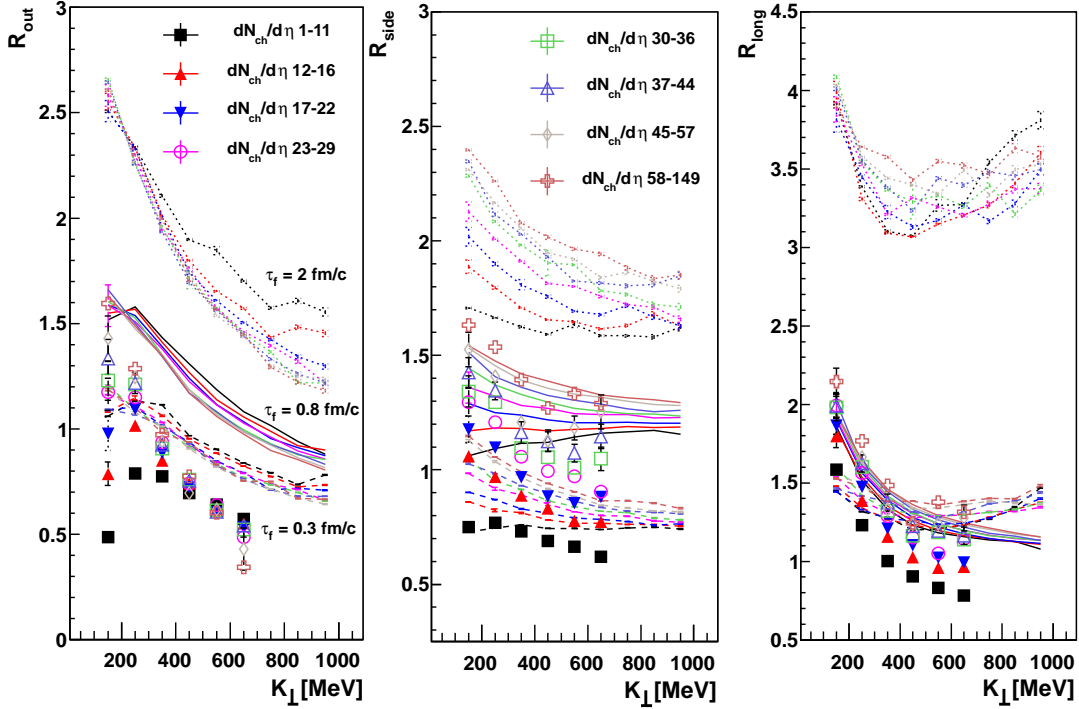}
     \caption{The lines represent HBT radii in p+p collisions at $\sqrt{s}$= 7 TeV from UrQMD for different
              multiplicities and formation times. The various line styles refer to results for $\tau_f=0.3$ fm/c
              (dashed), $\tau_f=0.8$ fm/c (default full lines) and $\tau_f=2$ fm/c (dotted). The colors
              represent the multiplicity classes as defined in table \ref{Tab:Multiplicities}. The points are
              data from the ALICE experiment \cite{Aamodt:2011kd}.}
     \label{Fig:RVsKt}
   \end{figure}

The $k_T$ dependence of the HBT radii extracted from the UrQMD freezeout
distribution is presented in Fig. \ref{Fig:RVsKt} for the $dN_{ch}/d\eta$
classes defined in table \ref{Tab:Multiplicities} in the pseudorapidity interval
$|\eta|<1.2$ in comparison to the ALICE data. The UrQMD calculations are
presented for different values of the formation time $\tau_f$ ($\tau_f$= 0.3
fm/c, dashed lines; $\tau_f$= 0.8 fm/c, full lines; $\tau_f$=2 fm/c, dotted
lines). For $\tau_f$=0.3 fm/c one obtains a good description for $R_{out}$,
while $R_{side}$ is slightly under predicted and the values for $R_{long}$ are
in line with data from ALICE. The choice $\tau_f$=0.8 fm/c leads to a slight
overestimation of $R_{out}$; however, it leads to a reasonable description of
$R_{side}$ data. Also the $k_T$ behaviour in $R_{long}$ is much closer to
the behaviour of the data. In contrast, a formation time of $\tau_f$=2 fm/c,
leads to a drastic overestimation of the data for all radii. Although
there are discrepancies between model and data for all values of $\tau_f$, the
sensitivity to $\tau_f$ is much larger than those discrepancies. Therefore, the
present ALICE data allows to constrain the formation time to values of $\tau_f
\approx$ 0.3-0.8 fm/c.\\

The overall shape of the radii as a function of
multiplicity and $k_T$ is also interesting. The $R_{side}$ radii (see Fig. \ref{Fig:RVsKt},
middle) from UrQMD and in the data are flat as a function of $k_T$ for low
multiplicity events. With increasing multiplicity, the radii develop a decrease
towards higher $k_T$. This is exactly the behavior one would expect for the
development of space-momentum correlations with rising event multiplicity
\cite{Werner:2011fd}. For $R_{out}$ (Fig. \ref{Fig:RVsKt}, left) however, there
is a $k_T$ dependence present in all multiplicity classes. Thus, only the
development of radial flow with rising particle multiplicity seems
insufficient to explain the $k_t$ dependence. Since $R_{out}$ and $R_{long}$
contain lifetime contributions of the source and $R_{side}$ does not, there
seem to be an additional non-trivial $k_t$ and multiplicity dependence in
the emission duration needed to explain the difference in the behavior of
$R_{out}$, $R_{side}$ and $R_{long}$. This additional correlation might be due
to an additional momentum dependence in $\tau_f$ apart from the trivial Lorentz
boost. This would lead to a direct effect on the emission duration, because it
changes the particles production spacetime points.

\subsection{Pb+Pb collisions at $\sqrt{s_{NN}}=2.76$ TeV}

\begin{figure}
\includegraphics[angle=0,width=0.9\textwidth]{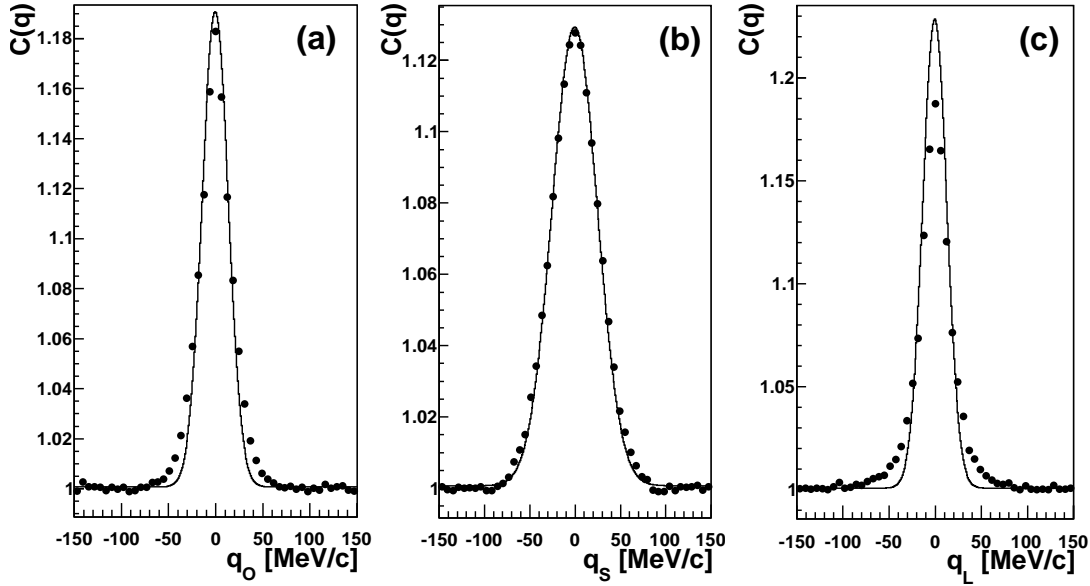}
\caption{Projections of the three-dimensional correlation function
(dots) and of the respective fit (lines) for the $k_T$ bin
$0.2-0.3$ GeV/c and $|\eta|<0.8$. When projecting on one axis the
other two components are restricted to the range (-30 $< q <$ 30)
MeV/c.} \label{pbpb-fig1}
\end{figure}

Firstly, the correlation functions are also calculated in bins of the transverse
momentum $k_T$. Fig.\ \ref{pbpb-fig1} shows the
projections of the three-dimensional correlation function (dots)
and of the respective fit (lines) for the $k_T$ bin $0.2-0.3$ GeV/c.
It is seen clearly that the correlator in sideward direction can be
described by a Gaussian form fairly well. However, it deviates
slightly from a Gaussian in the other two directions, especially in
the longitudinal direction, as found and discussed in previous
publications for HICs at lower energies \cite{Li:2008bk}. At LHC,
the fraction of excited unstable particles is much larger than at
lower beam energies, therefore, the non-Gaussian effect is more
pronounced in the current calculations. At RHIC energies, the
non-Gaussian effect was also seen in the experimental HBT
correlator, especially in the longitudinal direction
\cite{Adams:2004yc}. Furthermore, when comparing our fitting result
in Fig.\ \ref{pbpb-fig1} with that observed in experiment (in Fig. 1 of
Ref.\ \cite{Aamodt:2011mr}), it is seen that the non-Gaussian effect
is stronger in our calculations than in experiment. This might be
due to the omission of all potential interactions between particles
in the current cascade calculations. Support for this interpretation
was found in Ref.\ \cite{Li:2008bk}, where the consideration of a
mean-field potential plus Coulomb potential significantly reduced
the non-Gaussian effect on correlators of pions from HICs at AGS
energies. Therefore, both a dynamic treatment of the particle
transport with a proper EoS for the QGP phase
and the hadron phase, and further theoretical development of the
fitting formalism are equally important for a more precise
extraction of spatio-temporal information of the source
\cite{Pratt:2008bc}.

\begin{figure}
\includegraphics[angle=0,width=0.9\textwidth]{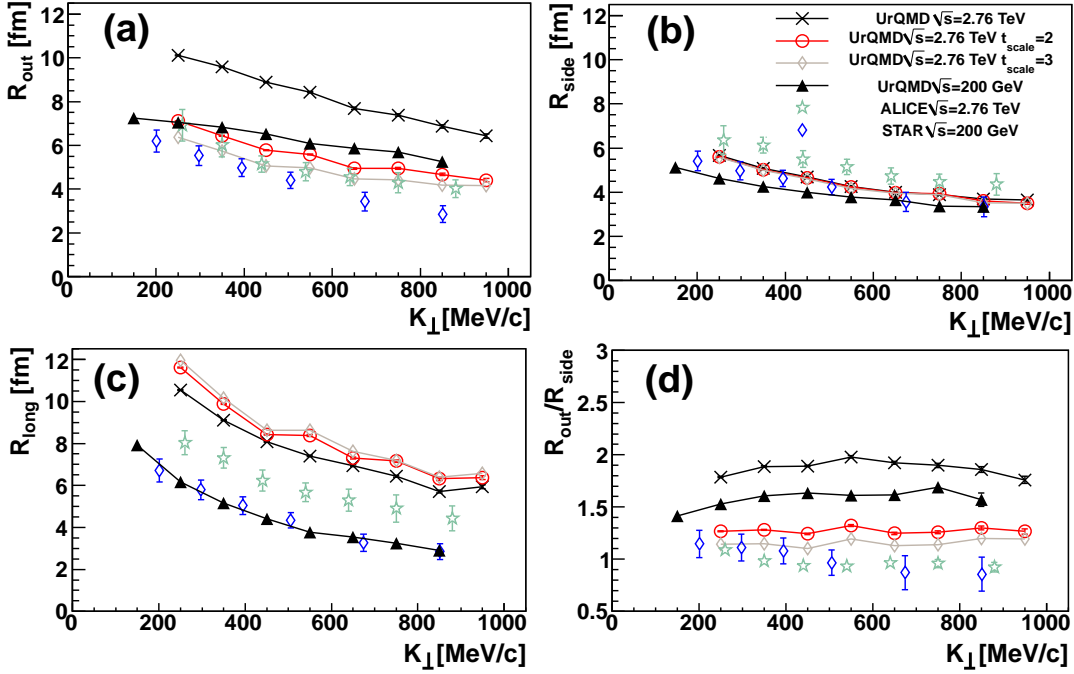}
\caption{$k_T$ dependence of pion HBT radii $R_{out}$
[panel (a)], $R_{side}$ [(b)], and $R_{long}$ [(c)], as well as the ratio
$R_{out}/R_{side}$ [(d)], for central ($\sigma/\sigma_T<5\%$) Pb+Pb
collisions at LHC energy $\sqrt{s_{NN}}=2.76$ TeV. For comparison,
parameters for central ($\sigma/\sigma_T<5\%$) Au+Au collisions at
RHIC energy $\sqrt{s_{NN}}=200$ GeV are also shown. Lines with
up-triangles and crosses are for model calculations while scattered
symbols are for experimental data of STAR/RHIC and ALICE/LHC
collaborations taken from \cite{Adams:2004yc,Aamodt:2011mr}. Lines
with circles and diamonds show results with an artificially
decreased emission duration by a factor of $t_{scale}=2$ and $3$,
separately, in the analysis of correlation function.} \label{pbpb-fig2}
\end{figure}

Besides the non-Gaussian effect, the contribution of the correlation
between the emission time and position to the HBT radii, especially
in the outward direction, has been paid more attention in recent
years since it closely relates to the stiffness of the EoS of
nuclear matter especially at the early stage of the whole dynamic
process \cite{Lin:2002gc,Li:2007yd,Li:2008qm}. It was found that, even in the
cascade calculation, there exists a visibly positive correlation
between the emission time and position \cite{Li:2012aa}. And, the most important contribution to
$R_{out}$ comes from the emission duration term. It implies that both a shorter
duration time and a stronger $x-t$ correlation lead to a smaller
$R_{out}$ value, which will be shown in Fig.\ \ref{pbpb-fig2}
specialized for the result of HBT radii.

Fig.\ \ref{pbpb-fig2} shows the $k_T$ dependence of the HBT radii $R_{out}$,
$R_{side}$, and $R_{long}$, as well as the ratio $R_{out}/R_{side}$, extracted from the
Gaussian fit to the two-pion correlators. The UrQMD cascade
calculations for central Pb+Pb collisions at LHC energy
$\sqrt{s_{NN}}=2.76$ TeV (lines with crosses) and central Au+Au
collisions at RHIC energy $\sqrt{s_{NN}}=200$ GeV (lines with
up-triangles) are compared to corresponding experimental data by
ALICE/LHC (open stars) and STAR/RHIC (open diamonds). A strong
decrease of the three HBT-radii with $k_T$ is seen both in
experiments and in the UrQMD calculations for HICs. This implies a
substantial expansion of the source and is qualitatively captured by
the UrQMD dynamics. Following experimental results, the calculated
HBT radii for Pb+Pb at LHC are found to be larger than those for
Au+Au at RHIC. The largest increase exists in the longitudinal
direction, which is also seen by the experiments. Although the
comparison of the calculated HBT radii $R_{long}$ and $R_{side}$ with
corresponding data at RHIC is fairly well, it gets worse when going
to the higher LHC energy. At LHC the calculated $R_{side}$ values at all
$k_T$ are found to be slightly smaller than data, while $R_{long}$ and
$R_{out}$ values are larger than data. Together with large calculated
$R_{out}$, the emission time related quantity $R_{out}/R_{side}$ is found to be
markedly larger than the data.

From Eq.~(\ref{eqn:RO}) it is clear that the
value of $R_{out}$ is strongly dependent on the emission duration of the
particles. To further investigate the contribution of the emission
duration to the HBT radii, we artificially decrease it by rescaling
the relative time $t$ to the ``effective source center time''
$\overline{t}$ ($=<t_i>$) by $t= t_i-\overline{t} \rightarrow
t'=(t_i-\overline{t})/t_{scale}$ in the calculation of the
correlation function at LHC energies. This effectively changes
Eq.~(\ref{eqn:RO}) to
\begin{equation}
 R_{out}'^2 = \langle (x-\beta_T t')^2 \rangle = \langle x^2 \rangle - 2 \frac{\langle \beta_T tx \rangle}{t_{scale}} + \frac{\langle \beta_T^2t^2\rangle}{t_{scale}^2}.
\end{equation}
The results for this calculation are presented as lines with circles
($t_{scale}=2$) and with diamonds ($t_{scale}=3$) in Fig.\
\ref{pbpb-fig2}. The artificially decreased emission duration leads to
smaller $R_{out}$ values in all $k_T$ bins but leaves $R_{side}$ unchanged,
as expected. Overall it results in an improved agreement with the
data of $R_{out}/R_{side}$ ratio. From Fig.~\ref{pbpb-fig2} it is also found that
$R_{long}$ is overestimated at LHC. Since $R_{long}$ is mainly related to the
lifetime of the source, it implies that this lifetime is also
overestimated by UrQMD. Other calculations in
\cite{Aamodt:2011mr,Graef:2012sh} show that UrQMD overestimates the
source lifetime by a factor of $\sim2-3$ when compared to LHC data.
The overestimation of both $R_{out}$ and $R_{long}$ can be attributed to the
known fact that the pressure in the early stage is not strong enough
in the cascade model calculations. A higher pressure would lead to a
more explosive expansion, a stronger phase-space correlation, and a
faster decoupling of the system, thus leading to smaller regions of
homogeneity. For more discussion we refer the reader to
\cite{Pratt:2008bc,Li:2007yd}. With the improved integrated
Boltzmann + hydrodynamics hybrid approach
\cite{Petersen:2008dd,Petersen:2011sb,Steinheimer:2010ib,Steinheimer:2011mp},
where various EoS of nuclear matter during the hydrodynamic
evolution may be treated consistently and a decoupling supplemented
by realistic three-dimensional hypersurfaces we hope to get a satisfactory solution
in the near future.

\section{Summary}
To summarize, two-pion HBT correlations for p+p and central Pb+Pb
collisions at the LHC energies are investigated with
the UrQMD model combined with a correlation afterburner CRAB. The transverse momentum dependence of the
Pratt-Bertsch HBT radii $R_{long}$, $R_{out}$, and $R_{side}$ is extracted from a three-dimensional
Gaussian fit to the correlator in the LCMS system. It is seen that the correlator in sideward direction can be
described by a Gaussian form fairly well while non-Gaussian effect is seen in both
longitudinal and outward directions.

In the p+p collisions, we have shown that the data provides rather
direct access to the particle formation times in the flux tube fragmentation.
The sensitivity to $\tau_f$ is large enough compared to the model uncertainties
to find that a value $\tau_f=0.3-0.8$ fm/c is strongly favored compared to
larger values for $\tau_f$. Values of $\tau_f \ge 2$ fm/c can be ruled out from
the present analysis.

In the Pb+Pb case, both the transverse momentum
$k_T$ dependence and the beam energy (from RHIC to LHC) dependence
of the HBT radii $R_{out}$, $R_{side}$, and $R_{long}$ exhibit qualitatively
the same behavior as found in the experiments. However, the
calculated $R_{out}/R_{side}$ ratios at all $k_T$ bins are found to be larger
than in the data, both at RHIC \& LHC. We traced this finding back
to the explosive dynamics of the fireball at LHC which results in
both a shorter emission duration and a stronger time-space
correlation than modeled here.

\ack{This work was supported by the Helmholtz International Center for FAIR
within the framework of the LOEWE program launched by the State of
Hesse, GSI, and BMBF. Q.L. thanks the financial support by the NNSF (Nos.
10905021, 10979023), the Zhejiang Provincial NSF (No. Y6090210), and
the Qian-Jiang Talents Project of Zhejiang Province (No. 2010R10102)
of China.}


\section*{References}
\begin{thebibliography}{9}

\bibitem{Adams:2004yc}
  Adams J {\it et al.}  [STAR Collaboration] 2005
  {\it Phys. Rev.}  C {\bf 71} 044906

\bibitem{Lisa:2000no}
Lisa M A {\it et al.} [E895 Collaboration] 2000 {\it Phys. Rev. Lett.} {\bf 84} 2798

\bibitem{Alt:2007uj}
Alt C {\it et al.} [NA49 Collaboration] 2008 {\it Phys. Rev.} C {\bf 77} 064908

\bibitem{Afanasiev:2002mx}
Afanasiev S V {\it et al.} [NA49 Collaboration] 2002 {\it Phys. Rev.} C {\bf 66} 054902

\bibitem{Adamova:2002wi}
Adamov\'{a} D {\it et al.} [CERES Collaboration] 2003 {\it Nucl. Phys.} A {\bf 714} 124

\bibitem{Abelev:2009tp}
Abelev B I {\it et al.} [STAR Collaboration] 2009 {\it Phys. Rev.} C {\bf 80} 024905

\bibitem{Back:2004ug}
Back B B {\it et al.} [PHOBOS Collaboration] 2006 {\it Phys. Rev.} C {\bf 73} 031901

\bibitem{Back:2005hs}
Back B B {\it et al.} [PHOBOS Collaboration] 2006 {\it Phys. Rev.} C {\bf 73} 021901

\bibitem{Back:2002wb}
Back B B {\it et al.} [PHOBOS Collaboration] 2003 {\it Phys. Rev. Lett.} {\bf 91} 052303

\bibitem{Abelev:2008ez}
Abelev B I {\it et al.} [STAR Collaboration] 2009 {\it Phys. Rev.} C {\bf 79} 034909

\bibitem{Lisa:2005dd}
Pratt S, Soltz R and Wiedemann U 2005
  {\it Ann. Rev. Nucl. Part. Sci.}  {\bf 55} 357

\bibitem{Aamodt:2011kd}
  Aamodt K {\it et al.}  [ALICE Collaboration] 2011
  {\it Phys. Rev.} D {\bf 84} 112004

\bibitem{Aamodt:2011mr}
  Aamodt K {\it et al.}  [ALICE Collaboration] 2011
  {\it Phys. Lett.}  B {\bf 696} 328

\bibitem{Karpenko:2011qn}
  Karpenko Y A and Sinyukov Y M 2011
  {\it J. Phys.} G {\bf 38} 124059

\bibitem{Humanic:2010su}
  Humanic T J 2011
  {\it arXiv:1011.0378 [nucl-th]}

\bibitem{Werner:2011fd}
  Werner K, Mikhailov K, Karpenko I and Pierog T 2011
  {\it arXiv:1104.2405 [hep-ph]}

\bibitem{Bass:1998ca}
  Bass S A {\it et al.} 1998
  {\it Prog. Part. Nucl. Phys.} {\bf 41} 255

\bibitem{Bleicher:1999xi}
  Bleicher M {\it et al.} 1999
  {\it J. Phys.} G {\bf 25} 1859

\bibitem{Petersen:2008kb}
  Petersen H, Bleicher M, Bass S A and Stocker H 2008
  {\it arXiv:0805.0567 [hep-ph]}

\bibitem{Petersen:2008dd}
  Petersen H, Steinheimer J, Burau G, Bleicher M and Stocker H 2008
  {\it Phys. Rev.} C {\bf 78} 044901

\bibitem{Li:2011aa}
Li Q F, Shen C W, Guo C C, Wang Y J, Li Z X, {\L}ukasik J and Trautmann W 2011
{\it Phys. Rev.} C {\bf 83} 044617

\bibitem{Sjostrand:2006za}
  Sjostrand T, Mrenna S and Skands P Z 2006
  {\it JHEP} {\bf 0605} 026

   \bibitem{Andersson:1983ia}
     Andersson B, Gustafson G, Ingelman G and Sjostrand T 1983
     {\it Phys. Rept.}  {\bf 97} 31

\bibitem{Mitrovski:2008hb}
  Mitrovski M, Schuster T, Graf G, Petersen H and Bleicher M 2009
  {\it Phys. Rev.}  C {\bf 79} 044901

\bibitem{Petersen:2011sb}
  Petersen H 2011
{\it Phys. Rev.}  C {\bf 84} 034912

\bibitem{Pratt:1994uf}
  Pratt S {\it et al.} 1994
  {\it Nucl. Phys.} A {\bf 566} 103C

\bibitem{Pratt:1986cc}
  Pratt S 1986
  {\it Phys. Rev.}  D {\bf 33} 1314

\bibitem{Bertsch:1988db}
  Bertsch G, Gong M and Tohyama M 1988
  {\it Phys. Rev.}  C {\bf 37} 1896


\bibitem{Wiedemann:1999qn}
  Wiedemann U A and Heinz U W 1999
  {\it Phys. Rept.}  {\bf 319} 145

\bibitem{Li:2008bk}
  Li Q and Bleicher M 2009
  {\it J. Phys.} G {\bf 36} 015111


\bibitem{Pratt:2008bc}
  Pratt S 2009
  {\it Acta Phys. Polon.}  {\bf B40} 1249

\bibitem{Lin:2002gc}
  Lin Z W, Ko C M and Pal S 2002
  {\it Phys. Rev. Lett.}  {\bf 89} 152301


\bibitem{Li:2007yd}
  Li Q, Bleicher M and Stocker H 2008
  {\it Phys. Lett.}  B {\bf 659} 525

\bibitem{Li:2008qm}
  Li Q, Steinheimer J, Petersen H, Bleicher M and Stocker H 2009
  {\it Phys. Lett.}  B {\bf 674} 111

\bibitem{Li:2012aa}
Li Q, Graef G and Bleicher M 2012
  {\it Phys. Rev.} C {\bf 85} 034908

\bibitem{Graef:2012sh}
  Graef G, Bleicher M and Li Q 2012
  {\it Phys. Rev.} C {\bf 85} 044901


\bibitem{Steinheimer:2010ib}
  Steinheimer J, Schramm S and Stocker H 2011
  {\it J. Phys.} G {\bf 38} 035001

\bibitem{Steinheimer:2011mp}
  Steinheimer J and Bleicher M 2011
 {\it Phys. Rev.} C {\bf 84} 024905


\end{thebibliography}
\end{document}